\documentclass[twoside,conference]{IEEEtran}
\IEEEoverridecommandlockouts
\usepackage{color,soul}

\usepackage{cite}
\ifCLASSINFOpdf
 \else
  \fi

\usepackage[cmex10]{amsmath}
\usepackage{amssymb}
\usepackage{mathtools}
\usepackage{amsthm}

\usepackage{graphicx}
\usepackage{bm}
\graphicspath{{figures/}}
\usepackage{epstopdf}
\begin{document}
\title{Polar Codes for SCMA Systems}
\author{Monirosharieh Vameghestahbanati, Ian Marsland, Ramy H. Gohary, and Halim Yanikomeroglu\\
Department of Systems and Computer Engineering, Carleton University, Ottawa, ON, Canada\\
Email: \{mvamegh, ianm, gohary, halim\}@sce.carleton.ca}
\date{}
\maketitle
\pagenumbering{gobble}
\begin{abstract}
   In this paper, we design and compare multilevel polar coding (MLPC) and bit-interleaved polar coded modulation (BIPCM) for uplink sparse code multiple access (SCMA) systems that operate over fast and block fading channels.
  Both successive cancellation (SC) and successive cancellation list (SCL) decoding algorithms are considered.
  Simulation results show that, with either decoder, BIPCM performs better than its MLPC counterpart. 
  Also, both BIPCM and MLPC exhibit a performance advantage over LTE turbo-coded and WiMAX LDPC SCMA systems when the SCL technique is used for decoding.



\end{abstract}
\section{Introduction}
\IEEEPARstart{T}{he} ever-increasing demand to accommodate various types of users and applications entails the emergence of novel techniques that are capable of coping with their requirements. The recently developed non-orthogonal multiple access (NOMA) waveform configuration of sparse code multiple access (SCMA) \cite{scma}
provides an innovative paradigm that can address the aforementioned demand \cite{huawei}.

In SCMA, sparse multidimensional codewords of multiple users are superimposed over shared orthogonal resources, whereby the number of users typically exceeds the number of resources. 
In essence, SCMA constitutes an instance of overloaded code division multiple access (CDMA) with two major differences: Firstly, in SCMA, the input binary data stream is directly encoded to multidimensional complex codewords chosen from a codebook set that is different for each user. In contrast, in CDMA the input data stream is mapped to QAM symbols followed by a CDMA spreader. Secondly, in SCMA the spreading matrix is restricted to be sparse such that only a few users overlap in each shared resource in order to minimize the multiuser interference; such a restriction is not imposed in CDMA. 

The performance of SCMA systems can be improved by channel coding, e.g., turbo and LDPC codes. 
In \cite{Sergienko16}, the bit error probability (BER) performance of turbo-coded SCMA systems was studied. A turbo principle with an iterative multiuser receiver for SCMA systems was proposed in \cite{wu15} and an uplink LDPC-coded SCMA system was studied in \cite{xiao15}.

Another class of channel codes is that of polar codes, which is based on channel polarization \cite{arikan}. Channel polarization involves synthesizing, out of $N$ independent binary discrete memoryless channel, a set of $N$ \emph{polarized} channels called the bit channels. Compared with other coding techniques, polar codes introduce an emerging type of error correcting code with the ability to achieve the capacity of discrete memoryless channels when the codeword length approaches infinity.
They also provide lower-complexity encoders and decoders compared with turbo and LDPC codes when decoded by successive cancellation (SC) \cite{niu14}. On the other hand, it is critical to construct polar codes to obtain the best performance when the codeword length is finite \cite{vangala15}. In \cite{Tal15}, successive cancellation list (SCL) decoding of polar codes was proposed to improve their performance at finite codeword lengths but with higher complexity compared to the SC decoding.

As both SCMA and polar coding are possible candidates for 5G systems \cite{huawei}, it is important to investigate the design of polar codes for SCMA systems from different aspects over different channel models.
A polar code design method for SCMA systems over additive white Gaussian noise (AWGN) was presented in \cite{Dai16_polar}.
In this paper, 
we design and compare the performance of two polar-coded schemes for uplink SCMA systems operating over two types of channels, fast and block fading. The first scheme is the so-called bit-interleaved polar coded modulation (BIPCM), whereas the other scheme is the so-called multilevel polar coding (MLPC). The underlying polar codes are designed using a Monte-Carlo simulation-based method , proposed in \cite{arikan}, that can be applied to different channels. However, instead of calculating the Bhattacharyya parameters of the bit channels as in \cite{arikan}, the BER is used \cite{vangala15}.

BIPCM is the combination of bit-interleaved coded modulation (BICM) with polar coding \cite{Afser14, mahdavi16}. In BIPCM, the message word for each user is encoded using a \emph{single} polar encoder to produce the output codeword. The codeword is then interleaved and passed to a signal mapper. A signal mapper maps the coded bits into points in the signal constellation with a certain labelling scheme. One such a labelling scheme is Gray labelling that generates bit levels that are independent from each other. 

MLPC is the combination of multilevel coding (MLC) with polar coding \cite{Seidl13}. In contrast with BIPCM, MLPC consists of \emph{multiple levels}; the number of levels corresponds to the number of bits in each symbol. At each level, each bit of the signal constellation is protected by an individual binary code. The total number of encoders and decoders in MLPC is equal to the number of bit levels. 
Each SCMA symbol depends on one bit from each encoder. 
In MLC, set partitioning (SP) labelling is known to perform better than other labelling scheme like Gray labelling \cite{MLC}, as it establishes larger variance in the reliability of each bit level. At the receiver, MLPC decodes the bit levels \emph{sequentially}, with information gained when decoding earlier levels used to make decisions at later levels. This is in contrast to BIPCM, where a single polar decoder is used to jointly decode the bits.

In this paper, we design and compare BIPCM and MLPC for SCMA systems that operate over block and fast fading channels.
  We employ multiuser detection and decoding separately in a concatenated manner as in \cite{scma} based on the non-binary message passing algorithm (MPA) detection, and both successive cancellation (SC) and successive cancellation list (SCL) decoding techniques. Simulation results show that, with either decoder, BIPCM performs better than its MLPC counterpart. Also, when the SCL technique is used for decoding, both BIPCM and MLPC exhibit a performance advantage over SCMA systems employing either the LTE turbo code or the WiMAX LDPC code.

The rest of the paper is organized as follows. We introduce the system model in Section \ref{sec:sysmodel}, and we present MPA detection for coded SCMA systems in Section \ref{sec:mpa}. In Section \ref{sec:pc}, we provide an overview on BIPCM and MLPC for SCMA systems. Section \ref{sec:sim} presents the simulation results, and Section \ref{sec:con} concludes the paper.
\begin{figure}[!t]
  \centering
  \includegraphics[width = 0.51\textwidth]{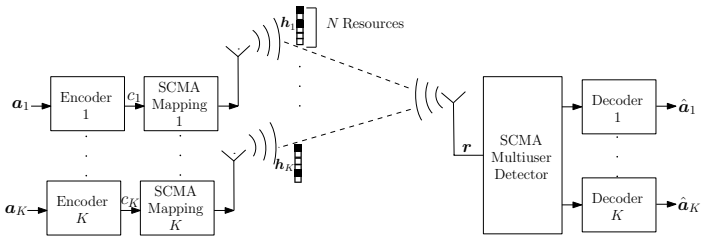}
 \caption{The system model for uplink transmission of a coded SCMA system with $K$ users multiplexed over $N$ resources with an overloading factor of $K/N$.}
\label{sysmodel}
\end{figure}
\section{System Model}\label{sec:sysmodel}
The system model is shown in Fig. \ref{sysmodel}. 
We assume there are $K$ users multiplexed over $N$ ($N<K$) orthogonal resources. In an $M$-ary signal constellation, each signal point represents $L_M=\log_2 M$ bits. Let $K_c$ and $N_c$ denote the length of message words input to the encoder and the length of codewords output from the encoder, respectively. The codewords are in turn partitioned into $N_c/L_M$ digital symbols of $L_M$ bits each. Let $c_k=\left(c_{k,1},\dots, c_{k, L_M}\right)$, $c_{k,m}\in \left(0,1\right)$, denote one symbol for user $k$, $k\in \left\{1,\dots, K\right\}$. Each symbol is mapped to an $N$-dimensional sparse complex codeword $\bm{x}_k=\left(x_{1,k}, \dots, x_{N,k}\right)^\textrm{T}$ selected from an $N$-dimensional complex codebook $\bm{V}_k$ of size $M$. That is, $x_{n,k}=V_{n,k}\left(c_k\right)$.

The receiver observes
  \begin{eqnarray}\label{r}
  \bm{r}=\sum\limits_{k=1}^{K}{\textrm{diag}}\left(\bm{h}_k\right) \bm{x}_k+\bm{w},
  \end{eqnarray}
where $\bm{h}_k=\left(h_{1,k}, \dots,h_{N,k}\right)^\textrm{T}$ denotes the fading channel coefficient vector for user $k$, and $\bm{w}~\thicksim\mathcal{CN}\left(0,\sigma^2 \mathbf{I}\right)$ is an $N$-dimensional complex Gaussian ambient noise sample with zero mean and a covariance matrix of $\sigma^2 \mathbf{I}$.

Due to the sparse nature of SCMA codewords, the near optimal MPA, which is discussed in Section \ref{sec:mpa}, is used to detect the SCMA codewords that will then be fed to the polar decoders to retrieve the message transmitted by each user. 
\section{Message Passing Algorithm}\label{sec:mpa}
 \begin{figure}[!t]
 \centering
\includegraphics[width = 0.21\textwidth]{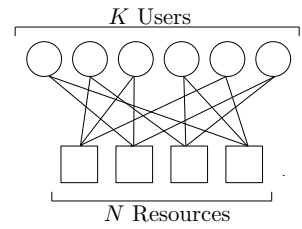}
 \caption{An example of the factor graph for $K=4$ users and $N=6$ resources.}
\label{fc}
\end{figure}
The structure of an SCMA code can be represented by an $N\times K$ binary mapping matrix $\bm{S}$ with its corresponding bipartite graph $\mathcal{G}\left(K,N\right)$ that contains $K$ variable nodes. 
and $N$ check nodes. 
The variable nodes and check nodes represent $K$ users and $N$ resources, respectively. Fig. \ref{fc} shows an example of $\mathcal{G}\left(K,N\right)$ with $K=6$, $N=4$, for the mapping matrix
\begin{eqnarray*}\label{S}
\bm{S} = \left[ {\begin{array}{*{20}c}
   0 & 1 & 1 & 0 & 1 & 0  \\
   1 & 0 & 1 & 0 & 0 & 1  \\
   0 & 1 & 0 & 1 & 0 & 1  \\
   1 & 0 & 0 & 1 & 1 & 0  \\
\end{array}} \right].
\end{eqnarray*}
 There is an edge between variable node $k$ and check node $n$ if and only if $s_{n,k}=1$. The set of check nodes connected to variable node $k$ is specified by the positions of the $1$'s in the $k$th column of the mapping matrix and is represented by $\mathcal{D}_k=\left\{n|s_{n,k}=1\right\}$.
~In a similar vein, the set of variable nodes connected to check node $n$ is identified by the positions of the $1$'s in the $n$th row of $\bm{S}$ and is denoted by $\mathcal{C}_n=\left\{k|s_{n,k}=1\right\}$.

Let $\bm{\mu}_{k\rightarrow n}$ and $\bm{\tilde{\mu}}_{n\rightarrow k}$ be vectors of length $M$ that represent the message passed from user node $k$ to resource node $n$, and the message from resource node $n$ to user node $k$ at each iteration, respectively.
Element $i$ of $\bm{\mu}_{k\rightarrow n}$ is given by 
\begin{eqnarray}\label{v2c}
  \bm{\mu}_{k\rightarrow n}\left(i\right) = \prod_{m \in \mathcal{D}_k \backslash n} \bm{\mu}_{m\rightarrow k}\left(i\right), 
\end{eqnarray}
where $i=1,\dots, M$.
Each element of the message conveyed from resource node $n$ to user node $k$ is
\begin{align}\label{c2v}
  \bm{\tilde{\mu}}_{n\rightarrow k}\left(i\right) = \sum_{\bm{c} \in \mathcal{M}^K|c_k=i} f\left\{r_n|\bm{c}\right\} \prod_{l \in \mathcal{C}_n \backslash k}\bm{\mu}_{l\rightarrow n}\left(c_l\right),
  \end{align}
  where $\bm{c}=\left(c_1,\dots, c_K\right)$ with elements $c_k \in \mathcal{M}=\left\{1,\dots, M\right\}$, and $r_n$ is the $n$th element of received vector $\bm{r}$. The likelihood function is given by
 \begin{eqnarray}\label{pdf}
    f\left(r_n|\bm{c}\right)=\frac{1}{2\pi\sigma^2}\exp\left\{-\frac{1}{2\sigma^2}\left|r_n-\sum_{k=1}^{K}h_{n,k}V_{n,k}\left(c_k\right)\right|^2\right\}.
  \end{eqnarray}
The \emph{a posteriori} probability distribution at each variable node is then
  \begin{equation}\label{p}
    \textrm{Pr}\left\{c_k=i|\bm{r}\right\}=\beta_{k}\prod_{m \in \mathcal{D}_k}\bm{\tilde{\mu}}_{m\rightarrow k}\left(i\right),
  \end{equation}
  where $\beta_{k}$ is chosen such that $\sum\nolimits_{i = 1}^M \textrm{Pr}\left\{c_k=i|\bm{r}\right\}=1$. 
\section{Polar codes for SCMA Systems}\label{sec:pc}
As mentioned earlier, we design and compare both BIPCM and MLPC for SCMA systems. In this section, we provide our design method with an overview on BIPCM and MLPC for SCMA systems. 
\vspace{-1 mm}
\subsection{Monte-Carlo Simulation-based Polar Code Design}
In general, the best polar code for one application is unlikely to be the best for another application, but it is possible to design good polar codes specifically for a certain application with a certain channel model. In contrast, turbo codes and LDPC codes cannot be specifically designed for each individual application, but turbo codes and LDPC codes designed for one application tends to be good for a wide variety of other applications.

There are several different polar code design methods in the literature e.g., \cite{arikan, tal13, Trifonov12}.
In this paper, we employ a Monte-Carlo simulation-based method, proposed in \cite{arikan}, that can be applied to different channels. However, instead of calculating the Bhattacharyya parameters of bit channels as in \cite{arikan}, the BER is directly estimated \cite{vangala15}. At a specific design SNR, we simulate the transmission of a large number of message words, and store the positions of the bits where the first error events occur\footnote{ The position of the first error event is the first message bit to be decoded incorrectly given that all previous bits are either decoded correctly or frozen. To speed up the design process, when an error occurs the position of the error is recorded, but a correct decision is fed back to the decoder, so that multiple first error events can be recorded with each simulated codeword.}. To achieve a required code rate  of $K_c/N_c$, the $K_c$ bit positions with the fewest recorded first error events are selected to carry the information bits. As will be discussed in Section \ref{sec:sim}, the performance of the system depends on the design SNR. 

\subsection{Bit-interleaved Polar Coded Modulation (BIPCM)}
In BIPCM, the message words for each user of length $K_c$ are encoded using a single polar encoder with a rate of $R$ to produce codewords of length $N_c$. The codewords are then interleaved using random interleaving, and partitioned into $L_M$-tuples to be mapped to complex codewords, $\bm{x}_k$.

At the receiver, the output symbol probabilities of the MPA i.e., $\textrm{Pr}\left\{c_k=i|\bm{r}\right\}$ given by (\ref{p}), are passed to the BIPCM decoder. Since $c_k$ can be expressed as $L_M$ bits, $c_k=\left(c_{k,1},\dots, c_{k, L_M}\right)$, we have $\textrm{Pr}\left\{c_k|\bm{r}\right\}=\textrm{Pr}\left\{c_{k,1}\dots c_{k, L_M}|\bm{r}\right\}$. The BIPCM decoder computes the log likelihood ratio (LLR) of each code bit as 
\begin{eqnarray}\label{llr1}
\lambda_{k,m}=\log\frac{\textrm{Pr}\left\{c_{k,m}=0|\bm{r}\right\}}
{\textrm{Pr}\left\{c_{k,m}=1|\bm{r}\right\}},
\end{eqnarray}
where $\lambda_{k,m}$ denotes the LLR corresponding to bit $m$, $m\in\left\{1,\hdots, L_M\right\}$, of user $k$, and
\begin{equation}\label{probBIPCM}
     \textrm{Pr}\left\{c_{k,m}=l|\bm{r}\right\} = \mathop{\sum_{c_{k,1}}\dots\sum_{c_{k,L_M}}}_{c_{k,m}=l}\textrm{Pr}\left\{c_{k,1}\dots c_{k, L_M}|\bm{r}\right\}
\end{equation}

\subsection{Multilevel Polar Coding (MLPC)}\label{sec:MLC} 
In the following, we describe the multilevel polar encoder and decoder schemes \cite{Seidl13} for SCMA systems.
\subsubsection{Multilevel Polar Encoder}
As mentioned earlier, for a signal constellation of $M$ points, there are $L_M=\log_2 M$ polar encoders, one for each bit.
Each polar encoder output has a length of $N_c/L_M$, so for each SCMA user we have a total of $N_c$ code bits. The code bits are fed to the mapper so that each symbol carries exactly one bit from each polar encoder.

In the design process of MLPC, all the constituent polar codes are designed simultaneously. That is, the first error event probabilities of all the possible $N_c$ message bits are calculated via Monte Carlo simulation, and then the best $K_c$ positions are selected, without regard to which constituent polar encoder they are associated with. As a result, the code rates of the constituent codes are determined automatically, and reflect the reliability of each bit level. In particular, those bit levels with lower reliability end up being assigned a higher rate code and vice versa. In keeping with polarization principle \cite{arikan}, we use set partitioning labelling to polarize the bit level reliabilities \cite{Seidl13}.
\subsubsection{Multilevel Polar Decoder}
The symbol probabilities $\textrm{Pr}\left\{c_k=i|\bm{r}\right\}$ produced by the MPA given by (\ref{p}) are passed to the multilevel polar decoder. Unlike BIPCM, with MLPC, the bit levels are decoded consecutively, with information gained when decoding earlier levels used to direct decision making at later levels. 
The probability distribution at the $m$th bit, given perfect knowledge of the bits at lower levels, $c_{k,1},\dots, c_{k,m-1}$, but no knowledge of the higher levels, $c_{k,m+1},\dots, c_{k,L_M}$, can be calculated as
\begin{eqnarray}\label{pmf}
\textrm{Pr}\left\{c_{k,m}|\bm{r}, c_{k,1},\dots, c_{k,m-1}\right\}= \hspace{3cm} \nonumber \\
\hspace{2.5cm}\sum_{c_{k,m+1}}\dots\sum_{c_{k,L_M}}
\frac{\textrm{Pr}\left\{c_{k,1},\dots, c_{k,L_M}|\bm{r}\right\}}{\textrm{Pr}\left\{c_{k,1},\dots, c_{k,m-1}|\bm{r}\right\}},
\end{eqnarray}
and the corresponding LLR is
\begin{equation}\label{llr}
\lambda_{k,m}=\log\frac{\textrm{Pr}\left\{c_{k,m}=0|\bm{r},c_{k,1},\dots,c_{k,m-1}\right\}}
{\textrm{Pr}\left\{c_{k,m}=1|\bm{r},c_{k,1},\dots,c_{k,m-1}\right\}}.
\end{equation}
Multilevel decoding is carried out by calculating the LLRs for the first level, $\lambda_{k,1}$, for all $N_c$ symbols for user $k$ and passing these LLRs to a polar decoder for the first-level polar code. This decoder provides an estimate of the transmitted polar codeword, which is used as $c_{k,1}$ when calculating the LLRs at the second level, $\lambda_{k,2}$. The process is repeated until all levels have been decoded.

 In this paper, we use both SC and SCL techniques for decoding BIPCM and MLPC. Note that the encoding complexity is in the order of $\mathcal{O}(N_c\log{}N_c)$ for BIPCM, and in the order of $\mathcal{O}(N_c\log{}N_c/L_M)$ for MLPC. In addition, to calculate the LLR corresponding to each bit level, \eqref{llr1} suggests that BIPCM requires $N_c$ summations. On the other hand, according to \eqref{llr}, the number of summations involved in the LLR calculation of level $m$ of MLPC is $N_c/\left(L_M\:2^{m-1}\right)$, which results in a total of at most $2N_c/L_M$ summations. As such, MLPC is computationally less complex compared with BIPCM. 
 \section{Simulation Results}\label{sec:sim}
 \begin{figure}[!t]
\centering
 \includegraphics[width = 0.48\textwidth]{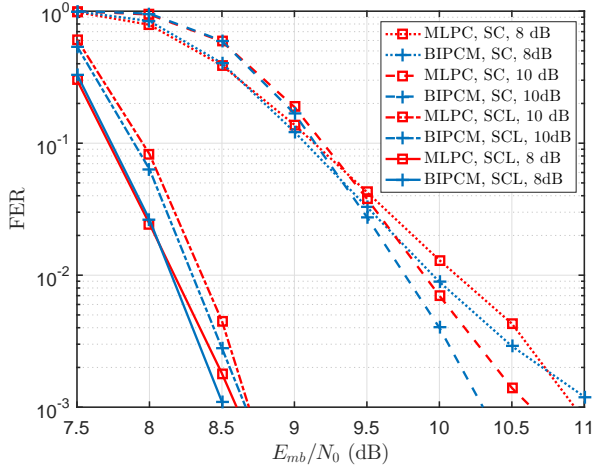}
 \caption{The comparison among an SC-decoded MLPC system, an SC-decoded BIPCM system, an SCL-decoded MLPC system, and an SCL-decoded BIPCM system, with $N_c = 2048$, at two different design SNRs. The overall code rate for all cases is $R = 2/3$. All systems operate in fast fading.}
\label{FER_FF_rate2third-snr}
\end{figure}
In this section, we evaluate the frame error rate (FER) performance of an uplink SCMA system with $K=6$ users and $N=4$ orthogonal resources with the widely used 4-dimensional complex codebooks of size $M=4$, $L_M = 2$, given in \cite{codebook}. We use Gray labelling for the BIPCM case 
and SP labelling for the MLPC case. Each user operates over a fading channel that is contaminated by AWGN, and transmits on only two of the four resources. We consider two fading scenarios: fast fading and block fading. In the fast fading scenario, the channel coefficients for each user vary independently in every use of the channel, while in the block fading case, the channel coefficients for each user are constant over a block of 18 channel uses as in \cite{Sergienko16}. In both channel models, we assume that each user uses the same channel coefficients over the $\left|\mathcal{D}_k\right|$ resources. 
Both BIPCM and MLPC when used by the SCL decoder, have a cyclic redundancy check (CRC) length of 16 bits and a list size of 32. The turbo codes are from the 3GPP LTE standard \cite{3gpp} and the LDPC code is chosen from the WiMAX standard \cite{wimax}.
\begin{figure}[!t]
\centering
 \includegraphics[width = 0.48\textwidth]{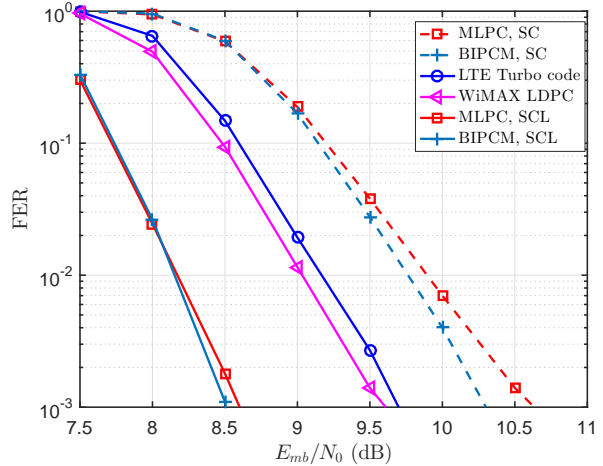}
 \caption{The comparison among an SC-decoded MLPC system, an SC-decoded BIPCM system, both with $N_c = 2048$, LTE turbo coded system with $N_c = 2070$, WiMAX LDPC system with $N_c=2016$, an SCL-decoded MLPC system, and an SCL-decoded BIPCM system, both with $N_c = 2048$. The overall code rate for all cases is $R = 2/3$. All systems operate in fast fading.}
\label{FER_FF_rate2third}
\end{figure}

Fig. \ref{FER_FF_rate2third-snr} compares the performance of length 2048 SC-decoded and SCL-decoded BIPCM SCMA systems, length 2048 SC-decoded and SCL-decoded MLPC\footnote{The length of the constituent polar code with MLPC for each level is 1024.} SCMA systems at two design SNRs, $E_{mb}/N_0 = 8$ dB and $E_{mb}/N_0 = 10$ dB. The net rate of all codes is $2/3$, and all systems operate in fast fading channel. We observe that it is important to choose the right design SNR depending on the target FER. In particular, to achieve a target FER of $10^{-2}$, a design SNR of $E_{mb}/N_0 = 10$ dB is a better choice for both BIPCM and MLPC with the SC decoder, while a design SNR of $E_{mb}/N_0 = 8$ dB is a better choice for both BIPCM and MLPC with the SCL decoder. 

In Fig. \ref{FER_FF_rate2third} we show the performance of length 2048 SC-decoded and SCL-decoded BIPCM SCMA systems designed at $E_{mb}/N_0 = 8$ dB, length 2048 SC-decoded and SCL-decoded MLPC systems designed at $E_{mb}/N_0 = 10$ dB, a length 2070 LTE turbo coded system, and a length 2016 WiMAX LDPC system. The net rate of all codes is $2/3$, and all systems operate in fast fading channel.
 As we see, BIPCM outperforms MLPC by about 0.2 dB with SC decoding, while it behaves marginally better than MLPC with SCL decoding.
Moreover, to achieve a target FER of $10^{-2}$, both BIPCM and MLPC with the SCL decoder outperform the LTE turbo coded system by about 1 dB, and outperform the WiMAX LDPC system by about 0.8 dB with a comparable decoding complexity. On the other hand, the LTE turbo and the WiMAX LDPC offer a gain of 0.5 dB and 0.7 dB compared with SC-decoded BIPCM system, respectively, but at a cost of much more decoding complexity.
 \begin{figure}[!t]
\centering
 \includegraphics[width = 0.46\textwidth]{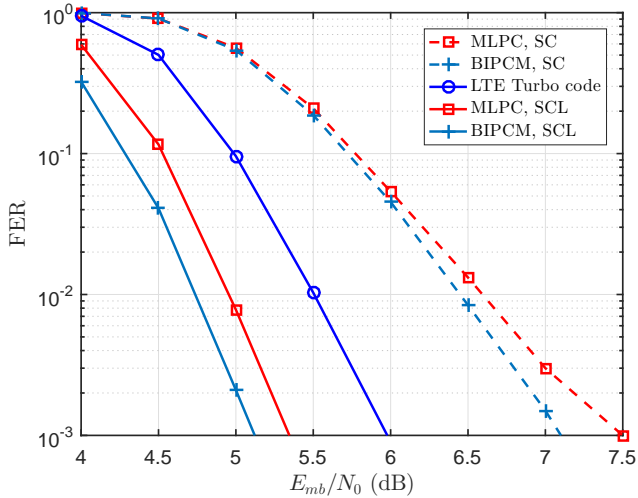}
 \caption{The comparison among an SC-decoded MLPC system, an SC-decoded BIPCM system, both with $N_c = 2048$, LTE turbo coded system with $N_c = 2028$, an SCL-decoded MLPC system, and an SCL-decoded BIPCM system, both with $N_c = 2048$. The overall code rate for all cases is $R = 1/3$. All systems operate in fast fading.} 
\label{FER_FF_ratethird}
\end{figure}

\begin{figure}[!t]
\centering
 \includegraphics[width = 0.46\textwidth]{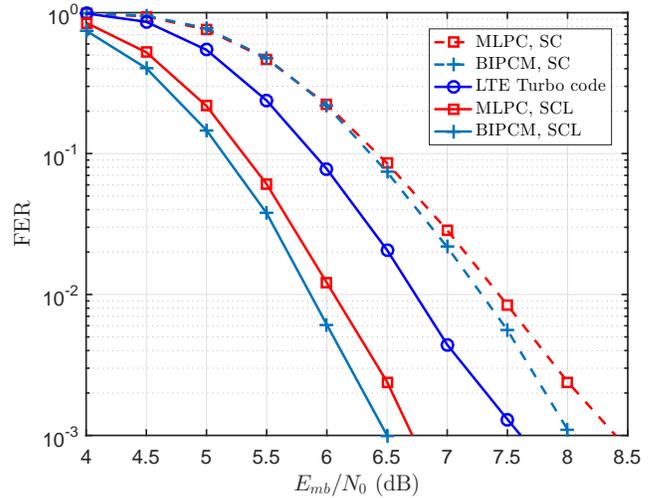}
 \caption{The performance comparison of an SC-decoded MLPC system, an SC-decoded BIPCM system, both with $N_c = 2048$, LTE turbo coded system with $N_c = 2028$, an SCL-decoded MLPC system, and an SCL-decoded BIPCM system, both with $N_c = 2048$. The net code rate for all scenarios is $R = 1/3$. All systems operate in block fading.}
\label{FER_RF_ratethird}
\end{figure}

We depict the performance comparison of rate $1/3$ coded SCMA systems that operate over fast fading and block fading channels in Fig. \ref{FER_FF_ratethird} and Fig. \ref{FER_RF_ratethird}, respectively. All polar codes are designed at $E_{mb}/N_0 = 6$ dB.
We see that MLPC is not an ideal candidate for SCMA systems. This is, however, in contrast with results previously reported in \cite{Seidl13} for single-user AWGN channels.
Particularly, we see that to achieve a target FER of around $10^{-2}$, the SC-decoded BIPCM system performs better than the SC-decoded MLPC system by about 0.2 dB in the fast fading case, and about 0.1 dB in the block fading scenario. The SCL-decoded MLPC results in a gain of about 0.25 dB in the case of fast fading and 0.15 dB in the case of block fading.
Moreover, Fig. \ref{FER_FF_ratethird} and Fig. \ref{FER_RF_ratethird} confirm that the performance of turbo coded system with $N_c=2028$ falls between BIPCM and MLPC with both SC decoding and SCL decoding. 
More precisely, the SCL-decoded BIPCM system performs better than the LTE turbo coded system by about 0.8 dB, in both fading scenarios, while the SC-decoded BIPCM system falls behind the LTE turbo coded system by about 0.9 dB in fast fading and about 0.5 dB in block fading.

\section{Conclusion}\label{sec:con}
We have designed and compared two polar-coded signalling schemes, MLPC and BIPCM, for uplink SCMA systems that operate over fast and block fading channels. The two polar-coded signalling schemes are designed with both SC and SCL decoding techniques. Simulation results show that, with either decoder, BIPCM performs better than MLPC. 
Moreover, both BIPCM and MLPC when decoded using SCL outperform the LTE turbo codes and the WiMAX LDPC. 
\section*{ACKNOWLEDGMENT}
This work is supported in part by an Ontario Trillium Scholarship, in part by Huawei Canada Co., Ltd., and in part by the Ontario Ministry of Economic Development and Innovations Ontario Research Fund - Research Excellence (ORF-RE) program.

\bibliographystyle{IEEEtran}
\bibliography{IEEEabrv,mybib}
\end{document}